\documentclass[floatfix,twocolumn,aps,prl,showpacs,amsmath,amssymb]{revtex4}
\usepackage[utf8]{inputenc}

\usepackage[dvips]{graphicx}
\usepackage{dcolumn}
\usepackage{bm}
\usepackage{dsfont}
\usepackage{color}
\usepackage{ulem} 
\usepackage{url}
\usepackage[colorlinks=true ,citecolor=blue]{hyperref}
\usepackage{amsmath,amssymb}
\usepackage{slashed}

\begin{document}

\title{ Two-dimensional Yukawa interaction driven by a nonlocal-Proca quantum electrodynamics}

\author{Van S\'ergio Alves$^{1}$, Tommaso Macrì$^{2,3,4}$, Gabriel C. Magalhães$^{1}$,  E. C. Marino$^5$, Leandro O. Nascimento$^{3,6}$}
\affiliation{$^1$ Faculdade de F\'\i sica, Universidade Federal do Par\'a, Av.~Augusto Correa 01, Bel\'em PA, 66075-110, Brazil  \\
$^2$ Departamento de Física Teórica e Experimental, Universidade Federal do Rio Grande do Norte, Natal-RN, Brazil\\ 
$^3$ International Institute of Physics, Campus Universit\' ario Lagoa Nova, C.P. 1613, Natal RN, 59078-970, Brazil  \\
$^4$ Dipartimento di Scienze Fisiche e Chimiche, Universit\`a dell’Aquila, via Vetoio, I-67010 Coppito-L’Aquila, Italy \\
$^5$ Instituto de F\'\i sica, Universidade Federal do Rio de Janeiro, C.P. 68528, Rio de Janeiro RJ, 21941-972, Brazil \\
$^6$ Faculdade de Ci\^encias Naturais, Universidade Federal do Par\'a, C.P. 68800-000, Breves, PA,  Brazil }

\date{\today}

\begin{abstract}

We derive two versions of an effective model to describe dynamical effects of the Yukawa interaction among 
Dirac electrons in the plane. Such short-range interaction is obtained by introducing a mass term for the intermediate 
particle, which may be either scalar or an abelian gauge field, both of them in (3+1) dimensions. Thereafter, we 
consider that the matter field propagates only in (2+1) dimensions, whereas the bosonic field is free to propagate
out of the plane. Within these assumptions, we apply a mechanism for dimensional reduction, which yields an 
effective model in (2+1) dimensions. In particular, for the gauge-field case, we use the Stueckelberg mechanism 
in order to preserve gauge invariance. We refer to this version as nonlocal-Proca quantum electrodynamics (NPQED). 
For both scalar and gauge cases, the effective models reproduce the usual $e^{-m r}/r$ Yukawa interaction in the 
static limit. By means of perturbation theory at one loop, we calculate the mass renormalization of the Dirac field. 
Our model is a generalization of Pseudoquantum electrodynamics (PQED), which is a gauge-field model that 
provides a Coulomb interaction for two-dimensional electrons. Possibilities of application to Fermi-Bose mixtures 
in mixed dimensions, using cold atoms, are briefly discussed.

\begin{center}
Subject Areas: Gauge field theories, Other nonperturbative techniques
\end{center}

\end{abstract}


\maketitle

\section{\textbf{I.\,Introduction}}

In the last decades, the interest of studying planar theories has increased in theoretical physics, mainly because 
of the discovery of new quantum effects, such as high-$T_c$ superconductivity, quantum Hall effect, and topological 
phase transitions \cite{Bernevig}. Furthermore, the emergence of both massless 
and massive Dirac excitations in two-dimensional materials, such as graphene \cite{grapexp} and silicene 
\cite{silexp}, has built a bridge between high-energy 
and condensed matter physics. For instance, well-known effects have been experimentally verified, or proposed 
in reachable energy scales, see \cite{Katsnelson} 
for an experimental realization of Klein paradox in graphene. On the other hand, for quantum chromodynamics 
the interest relies on the possibility of studying confinement in simplest models \cite{Appel}. 
More recently, ultracold atomic gases offered a clean and highly controllable platform for the quantum simulation of 
bosonic and fermionic systems \cite{Bloch}. Within such systems, static as well as dynamical properties of models in 
trapped geometries with short or long-range interactions \cite{Dalfovo99} can be probed, 
for example via their collective dynamics, or via density or momentum correlations 
\cite{Chiacchiera10}. 
Importantly, prototypical high energy physics models can be mapped into the low-energy, non-relativistic, many-body 
dynamics of ultracold atoms \cite{Zohar16}. Recently the experimental demonstration of a digital quantum simulation 
of the paradigmatic Schwinger model, a $U(1)$-Wilson lattice gauge theory \cite{Martinez16} was shown.

Among two-dimensional models, Pseudo quantum electrodynamics (PQED) \cite{marino,Kovner} 
(or Reduced quantum electrodynamics \cite{Teber}) has attracted some attention. This model describes the 
electromagnetic interaction in a system where electrons are confined to the plane, but photons 
(or the intermediating particle) may propagate out of the plane. Despite its nonlocal nature, PQED is still unitary 
\cite{unitarity}. Indeed, its main striking feature is that the effective action, for the matter field, remains the very same 
as the one provided by quantum electrodynamics in 3+1 dimensions, hence, unitarity is respected. 
In the static limit, it yields a Coulomb potential, which renormalizes the Fermi velocity in graphene \cite{Vozmediano}, 
as it has been verified by experimental findings \cite{Elias}. In the dynamical description, it is expected to generate 
a set of quantized-energy levels in graphene as well as an interaction driven quantum valley Hall effect \cite{PRX2015} 
at low enough temperatures. 
Furthermore, chiral-symmetry breaking has been shown to take place for both zero and finite temperatures 
\cite{VLWJF,CSBTemperature}. 
It also occurs in the presence of a Gross-Neveu interaction, whose main effect is to decrease the critical coupling 
constant, yielding a better scenario for dynamical mass generation \cite{PQEDGN}. Lower dimensional versions of PQED
have been investigated in Ref.~\cite{Tromb}, aiming for applications in cold atoms and in Ref.~\cite{CQED} for 
applications in the realm of topological insulators. 
All of these works rely on the fact that PQED generates a long-range interaction in the static limit, namely, the 
Coulomb potential $V_C(r)\propto 1/r$.   

In the meson theory of Yukawa, the so-called Yukawa potential $V(r)\propto e^{-m r}/r$ is 
a static solution of the motion equation $(-\nabla^2+m^2)V(r)=\delta(r)$, where $\delta(r)$ is the Dirac delta function 
\cite{Phystoday}. Within the quantum-field-theory interpretation, we may claim that the mediating field has a mass term. 
Motivated by this well-known result, we shall use the paradigm of including a mass term, for the intermediate particle, 
in order to generate a short-range interaction, i.e, the Yukawa potential in the plane. 
This potential has been applied to describe bound states \cite{Yukcrit3D,BSyukawa}, electron-ion interactions 
in a crystal \cite{Polaron}, and interactions between dark energy and dark matter \cite{abd} among others. 
Nevertheless, a planar quantum-field theory accounting for this interaction, for both static and dynamical 
regime, has not been derived yet.

In this paper, we show how one may include an interaction length in PQED, yielding a short-range and nonlocal theory. 
The simplest method is to generate a mass term for the mediating particle. Hence, we consider both the massive 
Klein-Gordon field as well as the massive Stueckelberg model. For both cases, we have a Yukawa 
potential between static charges. Thereafter, we calculate the mass renormalization at one loop in the small-coupling limit.

The outline of this paper is the following: in Sec.~II, we consider a Dirac field coupled to a scalar field, whose 
dynamics is given by the massive Klein Gordon model. In Sec.~III, we introduce the gauge field model. 
Since a naive addition of a mass for the gauge field would break gauge invariance, we consider the well known Stueckelberg action. 
This model is a generalized version of Proca quantum electrodynamics, on which we perform the dimensional reduction. 
In Sec.~IV, we compute the asymptotic behavior of the boson propagator at both small and large distances. 
In Sec.~V, we show that, by tuning the mass of the intermediate field, one may control the sign of the quantum 
correction, generated by the electron-self energy. We also include one appendix, where we present the details about 
the electron-self energy.

\section{\textbf{II.\,The Scalar Case}}

In this section, we perform a dimensional reduction of the Yukawa action in 3+1 dimensions. 
To generate an interaction length, we assume that the mediating particle is a massive real scalar field. 
Let us start with the Euclidean action in (3+1) dimensions, given by
\begin{equation}
{\cal L}_{\rm{4D}}= \frac{1}{2} \partial_\mu\varphi\partial^\mu\varphi+\frac{1}{2}m^2 \varphi^2 + 
g \varphi\bar\psi\psi +\bar\psi(i\partial\!\!\!/-M_0 )\psi, 
\label{action}
\end{equation}
where $g$ is a dimensionless coupling constant, $\varphi$ is a real and massive Klein-Gordon field, and $\psi$ is the Dirac field.

First, we calculate the effective action for the matter field ${\cal L}_{\rm{eff}}[\psi]$. In order to do so, 
we define the generating function $Z$
\begin{eqnarray}
Z&=&\int D\varphi D\bar\psi D\psi\exp\{-S[\psi,\varphi]\} \nonumber\\ 
&=&\int D\bar\psi D\psi D\varphi e^{\int d^4x [-\varphi(-\Box+m^2)\varphi/2-g\varphi\bar\psi\psi+O[\psi]]} \nonumber\\
&=&\int D\bar\psi D\psi \exp\{-S_{\rm{eff}}\}.
\label{Z}
\end{eqnarray}

Integrating out $\varphi$ in Eq.~(\ref{Z}) yields
\begin{equation}
{\cal L}_{\rm{eff}}[\psi]=-\frac{g^2}{2}\int d^4x d^4y(\bar\psi\psi)(x)\Delta_\varphi(x-y)(\bar\psi\psi)(y), \label{eff1}
\end{equation}
where
\begin{equation}
(-\Box+m^2)^{-1}\equiv\Delta_\varphi(x-y)=\int\frac{d^4 k}{(2\pi)^4}\frac{e^{-ik(x-y)}}{k^2+m^2}\label{phiprop}
\end{equation} 
is the free scalar-field propagator, which yields the interaction between the matter field. The static interaction $V(r)$ is 
provided by the Fourier transform of 
Eq.~(\ref{phiprop}) at $k_0=0$ (no time dependence), namely,
\begin{equation}
V(r)=\int\frac{d^3 k}{(2\pi)^3} \frac{e^{-i \textbf{k}.\textbf{r}}}{\textbf{k}^2+m^2}=\frac{e^{- m r}}{4\pi r} .  \label{Yukawa0}
\end{equation}
Eq.~(\ref{Yukawa0}) is the well-known Yukawa potential, where the inverse of $m$ is the interaction length of the model. 
This is just the consequence of the 
coupling $g\varphi\bar\psi\psi$ in (3+1)D. 

Next, we show how to generalize Eq.~(\ref{eff1}) and Eq.~(\ref{Yukawa0}) for 2+1 dimensions. Here, the main 
purpose is to keep the Yukawa potential between 
the matter field. In other words, we say that the fermionic field is confined to the plane, but the bosonic field is not. 
This is a roughly approximation of the derivation 
of PQED \cite{marino}. In order to do so, we assume that matter field is confined to the plane, i.e,
\begin{equation}
\bar\psi\psi(x)=\bar\psi\psi (x_0,x_1,x_2) \delta(x_3). \label{scalarcond}
\end{equation}
Using Eq.~(\ref{scalarcond}) in Eq.~(\ref{eff1}), we obtain
\begin{equation}
{\cal L}_{\rm{eff}}=-\frac{g^2}{2}\int d^3x d^3y(\bar\psi\psi)(x)G_\varphi(x-y)(\bar\psi\psi)(y), \label{eff2}
\end{equation}
where $G_{\varphi}(x-y)=\Delta_{\varphi}(x-y,x_3=0,y_3=0)$ is the effective scalar-field propagator in (2+1) dimensions, given by
\begin{equation}
G_{\varphi}(x-y)=\int\frac{d^3 k}{(2\pi)^3}e^{-ik(x-y)}\int\frac{dk_z}{(2\pi)} \frac{1}{k^2+k^2_z+m^2}. \label{prop0}
\end{equation}
Integrating over $k_z$ above, we find
\begin{equation}
G_{\varphi}(x-y)=\int\frac{d^3 k}{(2\pi)^3}\frac{e^{-ik(x-y)}}{2\sqrt{k^2+m^2}}. \label{scalarprop3D}
\end{equation}

In the static limit, Eq.~(\ref{scalarprop3D}) yields the Yukawa potential. Indeed,  
\begin{equation}
V(r)=\int\frac{d^2 k}{(2\pi)^2} \frac{e^{-i \textbf{k}.\textbf{r}}}{2\sqrt{\textbf{k}^2+m^2}}=\frac{e^{- m r}}{4\pi r},
\end{equation}
as expected from our dimensional reduction.

We may go beyond the static approximation by considering a nonlocal model with a propagator equal to Eq.~(\ref{scalarprop3D}). This is given by
\begin{equation}
{\cal L}_{3D}=\frac{1}{2}\partial^\mu\varphi K[\Box]\partial_\mu\varphi+\frac{1}{2}m^2\varphi^2+
g \varphi\bar\psi\psi +\bar\psi(i\partial\!\!\!/-M_0 )\psi, \label{scalar3D}
\end{equation}
with
\begin{equation}
K[\Box]\equiv \frac{2\sqrt{-\Box+m^2}}{-\Box}\equiv \int\frac{d^3 k}{(2\pi)^3}e^{ik x}\frac{2\sqrt{k^2+m^2}}{k^2}.
\end{equation}

From Eq.~(\ref{scalar3D}), it is straightforward that the scalar-field propagator is $G_{\varphi}$ and 
the effective action for the matter field are the very same 
as in Eq.~(\ref{scalarprop3D}) and Eq.~(\ref{eff2}), respectively. 

\section{ \textbf{III-The Gauge-Field Case}}

In this section, we consider that the matter field is coupled to a gauge field $A_\mu$, through a minimal 
coupling $A^\mu j_\mu$. The main steps are the same 
as in the previous calculation. Nevertheless, a massive term as $m^2 A_\mu A^\mu$ breaks gauge invariance. 
Hence, we must be careful about how to introduce 
the mass, i.e, the length scale for interactions. For the sake of simplicty, we consider an abelian field $A_\mu$, 
for which we may use the Stueckelberg mechanism. 
This is a mechanism for generating mass for $A_\mu$ without breaking gauge invariance \cite{Stuefield}. 
Before we perform the dimensional reduction, let us summarize this method.

First, we introduce a mass term into (3+1) QED, yielding the so-called Proca quantum electrodynamics,
whose action is given by
\begin{eqnarray}
{\cal L}_{\rm{4D}}&=& \frac{1}{4} F_{\mu \nu} F^{\mu\nu}+\frac{m^2}{2}A_\mu A^\mu +eA^\mu j_\mu \nonumber \\
&-&\frac{\lambda}{2}(\partial_\mu A^\mu)^2 +\bar\psi(i\partial\!\!\!/-M_0 )\psi, 
\label{action2}
\end{eqnarray}
where $e$ is the electric charge, $j_\mu=\bar\psi\gamma_\mu\psi$ is the matter current, $m^2$ is a 
massive parameter for the gauge field, and $\lambda$ is a 
gauge-fixing parameter. As expected,
\begin{equation}
\frac{m^2}{2}A_\mu A^\mu\rightarrow \frac{m^2}{2}A_\mu A^\mu+ m A^\mu \partial_\mu B +\frac{1}{2} (\partial_\mu B)^2 
\end{equation}
under $A_\mu\rightarrow A_\mu+\partial_\mu B/m$. Indeed, gauge invariance is explicitly broken.

Next, we introduce a scalar-field $B(x)$ (the Stueckelberg field) in Eq.~(\ref{action2}), hence,
\begin{eqnarray}
{\cal L}_{\rm{4D}}&=& \frac{1}{4} F_{\mu \nu} F^{\mu\nu}+\frac{m^2}{2}\left(A_\mu-\frac{\partial_\mu B}{m}\right)^2 +
eA^\mu j_\mu \nonumber \\
&-&\frac{\lambda}{2}(\partial_\mu A^\mu)^2 +\bar\psi(i\partial\!\!\!/-M_0 )\psi. 
\label{action3}
\end{eqnarray}
Eq.~(\ref{action3}) is known as Stueckelberg action. Despite the mass for the gauge field, it is invariant under 
gauge transformation, namely, $A_\mu\rightarrow A_\mu+\partial_\mu \Lambda$, $B\rightarrow B-m\Lambda$, 
and $\psi\rightarrow \exp(- i e \Lambda)\psi$ \cite{Stuefield}. Furthermore, it still produces the Yukawa 
interaction between static charges.

From Eq.~(\ref{action3}), we may find
\begin{eqnarray}
{\cal L}_{\rm{4D}}&=& \frac{1}{4} F_{\mu \nu} F^{\mu\nu}+\frac{m^2}{2}A_\mu A^\mu +\frac{(\partial_\mu B)^2}{2} \nonumber+ m A_\mu\partial^\mu B \\
&+&eA^\mu j_\mu -\frac{\lambda}{2}(\partial_\mu A^\mu)^2 +\bar\psi(i\partial\!\!\!/-M_0 )\psi. 
\label{action4}
\end{eqnarray}
We shall use Eq.~(\ref{action4}) to compute the effective action among the matter current $j_\mu$. 
The generating functional is
\begin{eqnarray}
Z&=&\int D\bar\psi D\psi DA_\mu D B\exp\{-S[\psi,A_\mu,B]\}.  
\end{eqnarray}
Integration over $B$ yields
\begin{eqnarray}
{\cal L}^{{\rm{eff}}}_{\rm{4D}}&=&\frac{1}{4} F_{\mu \nu} F^{\mu\nu}+\frac{m^2}{2}A_\mu 
A^\mu +\frac{\lambda_{\Box}}{2}(\partial_\mu A^\mu)^2 \nonumber\\
&+&eA^\mu j_\mu +\bar\psi(i\partial\!\!\!/-M_0 )\psi,  \label{effB}
\end{eqnarray}
where
\begin{equation}
\lambda_\Box=-\lambda+m^2\Box^{-1}.
\end{equation}

Now, for the sake of simplicity, we isolate the quadratic term in $A_\mu$, hence,
\begin{eqnarray}
{\cal L}^{{\rm{eff}}}_{\rm{4D}}&=&\frac{1}{2}A^\mu \left[(-\Box+m^2)\delta_{\mu\nu}+\lambda_\Box 
\partial_\mu\partial_\nu \right]A^\nu \nonumber\\
&+&eA^\mu j_\mu +\bar\psi(i\partial\!\!\!/-M_0 )\psi.  \label{effB}
\end{eqnarray}
Integrating out $A_\mu$, we get our desired effective action
\begin{equation}
{\cal L}^{\rm{eff}}_{{\rm{4D}}}[j]=-\frac{e^2}{2} j^{\mu}(x)\Delta_{\mu\nu} j^\nu(y) +\bar\psi(i\partial\!\!\!/-M_0 )\psi, \label{eff3}
\end{equation}
where
\begin{equation}
\Delta_{\mu\nu}=\frac{1}{-\Box+m^2}\left(\delta_{\mu\nu}-\frac{\lambda_\Box}{-\Box+m^2-\lambda_\Box}
\partial_\mu\partial_\nu\right). \label{propstue}
\end{equation}

Using Eq.~(\ref{propstue}) in Eq.~(\ref{eff3}) with charge conservation $\partial_\mu j^\mu=0$, 
we may conclude that, for a correct description, the gauge-field propagator is
\begin{equation}
\Delta_{\mu\nu}(x-y)=\int\frac{d^4 k}{(2\pi)^4}e^{-ik(x-y)}\frac{\delta_{\mu\nu}}{k^2+m^2}. \label{gaugeprop4D}
\end{equation}
In this way, all the gauge-dependence vanishes in the effective action.

In order to obtain the projected theory in (2+1) dimensions, we consider that the current matter 
only propagates in the plane, therefore,
\begin{eqnarray}
j^{\mu}(x) = \left \{ \begin{array}{ll}
                                              j^\mu (x_0,x_1,x_2)
                                              
                                            \delta(x_3), & \mu = 0, 1, 2,\\
                                                           0,                                      & \mu = 3.
                                      \end{array} \right . 
\label{constra}
\end{eqnarray}
Similarly to the previous case, this shall lead to 
\begin{equation}
G_{0,\mu\nu}=\int\frac{d^3 k}{(2\pi)^3}e^{-ik(x-y)}\frac{\delta_{\mu\nu}}{2\sqrt{k^2+m^2}}, \label{gaugeprop3D}
\end{equation}
which is the effective propagator in (2+1) dimensions.

Our main goal is to find the corresponding theory in (2+1) dimensions with the same effective action in Eq.~(\ref{eff3}). This model reads
\begin{eqnarray}
{\cal L}_{\rm{3D}}&=& \frac{1}{2} F_{\mu \nu} K[\Box] F^{\mu\nu}+\lambda A^\mu \partial_\mu K[\Box]\partial_\nu A^\nu +eA^\mu j_\mu \nonumber \\
&+&\bar\psi(i\partial\!\!\!/-M_0 )\psi, \label{PPQED}
\end{eqnarray}
where $\lambda$ is a gauge-fixing parameter. It is straightforward to show that, after integrating out 
$A_\mu$ in Eq.~(\ref{PPQED}), we obtain the same effective action as in Eq.~(\ref{eff3}) with the constraint 
in Eq.~(\ref{constra}). For an arbitrary $\lambda$, the free gauge-field propagator reads 
\begin{equation}
G_{0,\mu\nu}(k)=\frac{1}{2\sqrt{k^2+m^2}}\left[\delta_{\mu\nu}+\frac{(1-\lambda)}{\lambda}\frac{k_\mu k_\nu}{k^2}\right].  \label{gaugepropagator}
\end{equation}

There are two main features of Eq.~(\ref{PPQED}): (a) The massive parameter $m$ is no longer a pole 
of the gauge-field propagator, hence, it can not be thought as a mass and (b) Gauge-invariance is explicitly 
respected, i.e, there is no need to deal with Stueckelberg fields. Indeed, we could set $B=0$ from the very 
beginning, which means starting with Proca quantum electrodynamics and, therefore, breaking of gauge 
invariance. Then, after dimensional reduction, the corresponding 3D theory is still the same as in Eq.~(\ref{PPQED}) 
and that it is gauge invariant.

\section{ \textbf{IV- Asymptotic Behavior of Both Scalar and Gauge-field Propagators}}

In this section, we calculate the asymptotic expressions of $G_\varphi(x-y;m)$ and $\delta^{\mu\nu}G_{0\mu\nu}(x-y;m)$, i.e, propagators in the space-time coordinates. We use the integral version in Eq.~(\ref{prop0}), which has an extra $k_z$-integral. Hence,
\begin{equation}
G_{\varphi}(x-y;m)=\int\frac{d^3 k}{(2\pi)^3}e^{-ik(x-y)}\int_{-\infty}^{+\infty}\frac{d\mu}{(2\pi)} \frac{1}{k^2+\mu^2+m^2}. \label{propasyp0}
\end{equation}
Note that we have replaced $k_z$ by $\mu$, since $\mu$ is just a parametric variable. Thereafter, we use Eq.~(\ref{Yukawa0}) for solving the $k$-sphere integral, therefore,
\begin{equation}
G_{\varphi}(x-y;m)=\int_{-\infty}^{+\infty}\frac{d\mu}{(2\pi)} \frac{1}{4\pi |x-y|} e^{-\sqrt{(\mu^2+m^2)} |x-y|}. \label{propasyp1}
\end{equation}

Next, after solving the $\mu$-integral (see Ref.~\cite{Grads}), we have
\begin{equation}
G_{\varphi}(x-y;m)=\frac{m}{4\pi^2 |x-y|} K_1(m |x-y|), \label{Gxy}
\end{equation}
where $K_1$ is a modified Bessel function of the second kind. In the short-range limit $m |x-y|\ll 1$, it yields
\begin{equation}
G_{\varphi}(x-y;m)\approx \frac{1}{4\pi^2|x-y|^2},
\end{equation}
whereas in the long-range limit $m |x-y|\gg 1$, we find
\begin{equation}
G_{\varphi}(x-y;m)\approx\sqrt{\frac{m\pi}{2}} \frac{e^{-m |x-y|}}{4\pi^2|x-y|^{3/2}}.
\end{equation}
The gauge-field propagator may be calculated by following the very same steps.

We have described some general features of NPQED, in particular, its derivation, two-point functions, and interactions. Next, we shall explore quantum corrections by using perturbation theory.

\section{ \textbf{V- Perturbation Theory Results}}

In this section, we calculate the renormalized electron mass $m_R$ of the model in Eq.~(\ref{PPQED}) at one-loop approximation. The details about the calculation are shown in Appendix A. In particular, we would like to obtain its dependence on $m$, the mass term of the gauge field. Note that in our 3D model, this parameter must to be understood as the inverse of the interaction length. This, nevertheless, is the mass of the intermediate field that propagates in 4D. Thereafter a standard calculation, we obtain
\begin{equation}
z_R=1+\frac{\alpha}{2\pi} f(z),  \label{zR}
\end{equation}
where $z_R\equiv m_R/M_0$, $z\equiv m/M_0$, and
\begin{equation}
f(z)=\int_0^1dx \frac{2+x}{\sqrt{1-x}}\ln\left[\sqrt{(1-x)(z^2-x)+x}\right]. \label{fxz}
\end{equation}

\begin{figure}[h!]
\centering
\includegraphics[width=1.0\columnwidth]{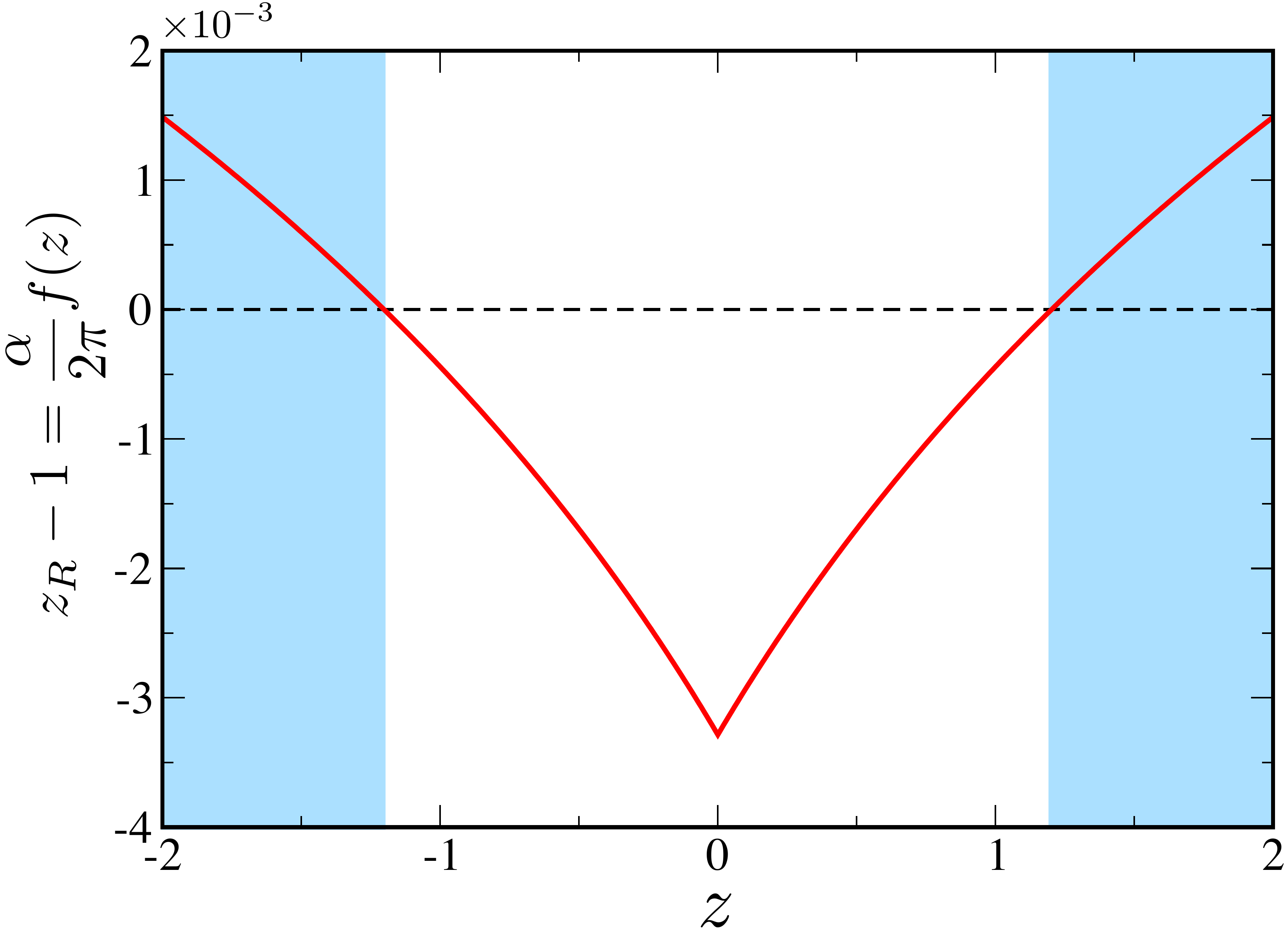}
\caption{{\it Quantum correction for the electron mass.}
We plot the quantity $\alpha f(z)/2\pi$ in Eq.~(\ref{zR}), where $z$ is the ratio between the 
inverse of interaction length $m$ and the bare electron mass $M_0$. 
For $|z|\leq z_c\approx 1.2$, we have $f(z)<0$, 
therefore, the renormalized energy gap $\delta_R E$ is less than the bare energy gap $2M_0$. 
On the other hand, $|z|\geq z_c$ yields $f(z)> 0$ (shaded area), i.e, the 
renormalized energy gap is larger than the bare energy gap. Finally, for $|z|=z_c$, 
we find $f(z_c)=0$ and the renormalized energy gap is equal to $2M_0$.} \label{figdia}
\label{fig3}
\end{figure}

From Eq.~(\ref{fxz}), it is clear that $f(z)=f(-z)$, therefore, the corrections are only dependent on the modulus of $m$. From now on, we assume $\alpha=1/137$. Next, we would like to address the effects of the $m$ parameter on $z_R$. Surprisingly, for $|z|\leq z_c\approx 1.2$, the quantum correction $\alpha f(z)/2\pi$ is negative, while for $|z|\geq z_c$, they become positive and cross the free-energy level $M_0$, see Fig.~\ref{figdia}.

Let us calculate the energy gap of the renormalized state 
$\delta_R E= E^+_R- E^-_R=M_0+ M_0\alpha f_x(z)/2\pi-(-M_0-M_0\alpha f(z)/2\pi)=2M_0+M_0\alpha f(z)/\pi.$, 
where $E^+_R>0$ and $E^-_R<0$ are the positive and negative energies, see  Appendix A. 
Since $f(z)$ may be negative for $|z|\leq z_c\approx 1.2$, 
we have that $\delta_R E$ lies inside the energy gap $2M_0$. 
The case $z\leq z_c$ resembles the quantized energy levels calculated in Ref.~\cite{PRX2015} or Ref.~\cite{PQEDGN}, 
because that renormalized states are closer to the zero-energy level. Nevertheless, because we are in the small 
coupling limit, hence, we do not generate dynamical symmetry breaking, since $z_R\rightarrow 0$ when $M_0\rightarrow 0$.

\section{ \textbf{VII.\, Discussion}}

PQED has been applied to describe the interactions of two-dimensional electrons, in particular, 
graphene in the strong-coupling regime. The main results rely on the fact that electrons do interact trough 
the Coulomb potential, which has an infinite range. Here, nevertheless, we consider 
a scenario where interactions have a finite range. We have applied the very same procedure for 
deriving PQED \cite{marino}, but considering a massive photon in (3+1) dimensions. 
Our main result is that, after performing the dimensional reduction, one obtains that the matter field 
interacts through a Yukawa potential in the static limit. Similar to PQED, it yields a nonlocal theory in both space 
and time.  Since the matter field is not relevant for the dimensional reduction, this model may be generalized to 
describe the Yukawa interaction between other kind of particles. 
Although our derivation follows standard steps of QED literature, we believe that they 
shall be relevant, in particular, for applications in condensed matter physics and cold-atom systems.

In Ref.~\cite{Monica}, the authors proposed a realization of a cold-atom system made of fermions in 
two-dimensions with bosons in three-dimensions. 
This has been called Fermi-Bose mixture in mixed dimensions. Accordingly to their theoretical model, the static 
interaction between the fermionic particles is given by the Yukawa potential. Although, in their approach, 
retardation effects are neglected, we may make use of NPQED for investigating these new regime. 
Indeed, besides anomalies \cite{PRX2015}, it has been shown that a full dynamical description provides new 
results for PQED in application to exciton spectrum in transition-metal dichalcogenides \cite{TMDPQED}.

We finally discuss the importance of the order in which the following two operations are implemented, namely: a) the mass generation for the gauge field; b) the dimensional reduction.  Interestingly, the result is sensitive to the order in which the inclusion of a mass and the 
dimensional reduction are performed. As a matter of fact, by doing ``a" before ``b", we have shown that the Yukawa potential $V(r)= e^{-mr}/4 \pi r$ 
is obtained from the static limit of the gauge-field propagator, given by Eq.~(\ref{gaugepropagator}). This is proportional 
to $1/(2\sqrt{p^2+m^2})$. 

Conversely, let us start with a massless gauge field, hence, by applying the dimensional reduction, 
we arrive at the PQED model \cite{marino}, whose propagator is proportional to $1/(2\sqrt{p^2})$. Suppose we now couple the PQED 
model, to a Higgs field in the broken phase, such that a massive term is generated to the gauge field. This changes the 
propagator from $1/(2\sqrt{p^2})$ to $1/(2\sqrt{p^2}+m)$, which clearly shows that we shall obtain a different model 
from the one associated to Eq.~(\ref{PPQED}). The potential is now given by a 
combination of Coulomb and Keldysh \cite{K} potentials, i.e, $V(r)\propto 1/r-m \left [\textbf{H}_0(mr)- Y_0(mr)\right ]$, 
where $\textbf{H}_0(mr)$ and 
$Y_0(mr)$ are Struve and Bessel functions, respectively. This potential does not decay exponentially at large distances as the Yukawa potential, rather, it has a power-law decay. We conclude, therefore, that we must to be careful about how one wishes to use quantum electrodynamics in applications for lower-dimensional systems, because of this sensitive relation between the dimensional reduction and phenomenological parameters, such as a mass for the gauge field.

\section{\textbf{acknowledgments}}

We are grateful to Rodrigo Pereira for very interesting and stimulating discussions. 
V. S. A. and G. C. M. acknowledges CAPES for financial support. 
T. M. acknowledges CNPq for support
through Bolsa de produtividade em Pesquisa n.
311079/2015-6. 
L. O. N. acknowledges the financial support from MEC/UFRN. 
E.C.M. is partially supported by CNPq and FAPERJ.

\textbf{\section{\textbf{Appendix A: Electron-self Energy: Isotropic Case}}}

In this Appendix, we show some details about the electron-self energy. First, let us write the Feynman rules of Eq.~(\ref{PPQED}) in Euclidean space. 
The free electron-propagator reads
\begin{equation}
S_{0F}=\frac{1}{\slashed{p}-M_0}=\frac{-(\slashed{p}+M_0)}{p^2+M^2_0}, \label{diracfree}
\end{equation} 
the gauge-field propagator (in the Feynman gauge $\lambda=1$) is
\begin{equation}
G_{0,\mu\nu}(p)=\frac{\delta_{\mu\nu}}{2\sqrt{p^2+m^2}},
\end{equation}
and the vertex interaction is $\Gamma^\mu=\gamma^\mu e$. The Euclidean matrices satisfy $\{\gamma_\mu,\gamma_\nu\}=-2\delta_{\mu\nu}$. 
From Eq.~(\ref{diracfree}), we have that the pole of the free Dirac electron is $p^2_E=-M_0^2$, where $p^2_E$ is the Euclidean momentum. 
For calculating the physical mass, one must to return to the Minkowski space, using $p^2_E\rightarrow -p^2_M$, such that $p^2_M=M^2_0$ are the physical poles.

The corrected electron propagator $S_F$ is
\begin{equation}
S_F=S_{0F}+S_{0F} (\Sigma) S_{0F}+...
\end{equation}
therefore,
\begin{equation}
S^{-1}_F=S_{0F}^{-1}- \Sigma(p). \label{fullelecprop}
\end{equation}
The electron-self energy reads
\begin{equation}
\Sigma(p)=e^2\int \frac{d^3k}{(2\pi)^3}\Gamma^\mu S_{0F}(k)\Gamma^\nu G_{0,\mu\nu}(p-k).  \label{ESE}
\end{equation}

Eq.~(\ref{ESE}) has a linear divergence, therefore, we need to use a regularization scheme. We choose to use the usual dimensional regularization $\epsilon=3-D$, where $D$ is an arbitrary dimension, which we shall consider $D\rightarrow 3$ in the very end of the calculation. After application of standard methods, we find
\begin{equation}
 \Sigma(p)= 2 C I+ 2C \left(\ln \frac{\mu}{M_0}+\frac{1}{\epsilon}\right) \left(\frac{2 \slashed{p}}{3}-6 M_0\right), \label{ESE2}
\end{equation}
where $\mu$ is an arbitrary massive parameter, generated by the prescription $e\rightarrow e \mu^\epsilon$, with $\epsilon=D-3$, where $D$ is the dimension of the space-time. This is a standard step in the dimensional regularization. The constant $C$ is given by
\begin{equation} \label{eq_C}
C=\frac{e^2}{16 \pi^2}
\end{equation}
and the parametric integral reads
\begin{equation}
I=\int_0^1 dx \frac{\slashed{p}(x-1)+ 3 M_0}{\sqrt{1-x}}\ln \frac{\Delta}{M_0},
\end{equation}
with
\begin{equation}
\Delta=\sqrt{p^2 x(1-x)+M^2_0 x+m^2(1-x)}.
\end{equation}

Eq.~(\ref{ESE2}) has both finite and a regulator-dependent terms, which must be eliminated by some renormalization scheme. For the sake of simplicity, we choose the minimal subtraction procedure, which essentially avoid the poles by introducing counter-terms in the original action. Hence,
\begin{equation}
\Sigma^R(p)=\lim_{\mu,\epsilon \rightarrow 0} (\Sigma(p,\mu,\epsilon)-CT)= A(p)\slashed{p}+B(p), \label{sigR}
\end{equation}
where CT stands for counter-terms,
\begin{equation}
A(p)= 2 C\int_0^1 dx \frac{(x-1)}{\sqrt{1-x}}\ln\frac{\Delta}{M_0}, \label{Ap}
\end{equation}
and
\begin{equation}
B(p)=  2 C \int_0^1 dx \frac{3 M_0  }{\sqrt{1-x}}\ln\frac{\Delta}{M_0}.  \label{Bp}
\end{equation}

Using Eq.~(\ref{sigR}) in Eq.~(\ref{fullelecprop}), we find
\begin{equation}
S_F^{-1}(p)=[(1-A(p))\slashed{p}-(M_0+B(p))]. \label{fue2}
\end{equation}

\begin{figure}[h!]
\centering
\includegraphics[width=1.0\columnwidth]{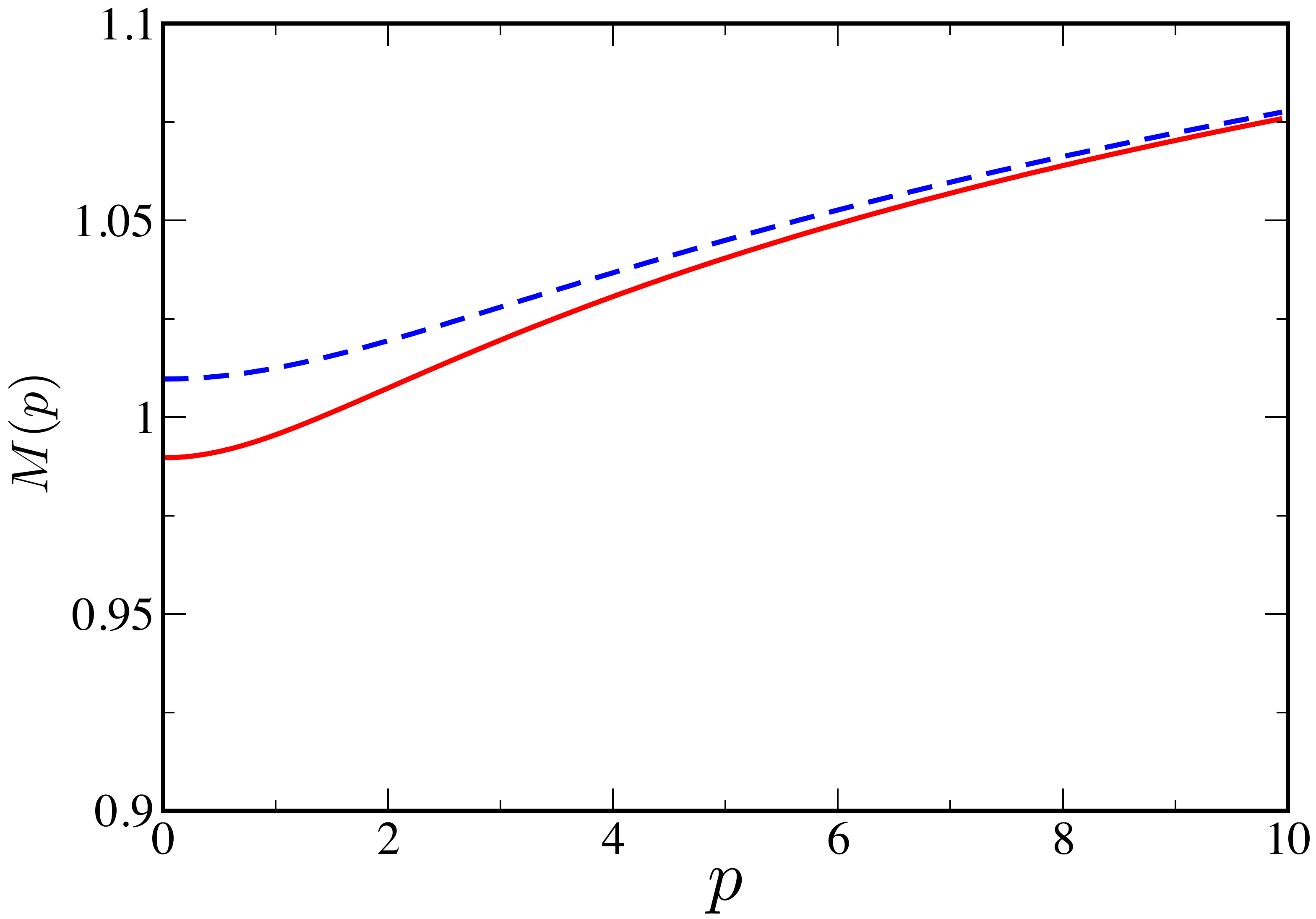}
\caption{{\it Mass function $M(p)$ in Eq.~(\ref{Mp}).}
For this plot we set $M_0=1$ and $\alpha=1/137$. Dashed line: $m/M_0=1.5$, full line: $m/M_0=0.5$. 
Note that $M(p=0)>M_0$ for the dashed line, but $M(p=0)<M_0$ for the full line.} 
\label{fig2}
\end{figure}

Multiplying Eq.~(\ref{fue2}) by $(1+A(p))$, we have
\begin{equation}
(1+A(p))S_F^{-1}(p)=-i[\slashed{p}-(M_0+B(p))(1+A(p))].
\end{equation}
Next, we define the renormalized matter field $\psi_R$, namely,
\begin{equation}
\psi_R=Z^{1/2}_\psi \psi=\sqrt{1-A(p)}\psi.
\end{equation}
Therefore,
The physical propagator reads
\begin{equation}
S^{-1}_{RF}=\frac{1}{\slashed{p}-M(p)}=\frac{-[\slashed{p}+M(p)]}{p^2+M^2(p)},  \label{srf}
\end{equation}
with
\begin{equation}
M(p)=M_0+B(p)+M_0 A(p). \label{Mp}
\end{equation}
Using Eqs.~(\ref{eq_C}), (\ref{Bp}), (\ref{Ap}) in Eq.~(\ref{Mp}), we find
\begin{equation}
M(p)=M_0+\frac{\alpha}{2\pi} M_0 \int_0^1 dx\frac{(2+x)}{\sqrt{1-x}}\ln\frac{\Delta}{M_0}. \label{Mpfinal}
\end{equation}
Eq.~(\ref{Mp}) yields the so-called mass function $M(p)$, which is, essentially, the momentum-dependent part of the 
electron-self energy that renormalizes the electron mass \cite{Appelquist1985}. In Fig.~\ref{fig2}, we show that by 
using different values of $m/M_0$, we may generate quantum corrections that either increase or decrease the 
renormalized mass in comparison with the bare value $M_0$. To clarify this result, we shall calculate the renormalized 
mass $m_R$.

The pole of Eq.~(\ref{srf}) is given by the solution of $p^2_E=- M^2(p^2_E)$, which, in the Minkowski space, 
yields $m^2_R=M^2(-m^2_R)$, where $m^2_R$ is the renormalized mass. Note that we have applied 
$p^2_E\rightarrow -m^2_R$ for calculating the pole in the Minkowski space. Furthermore, because we are 
at one-loop approximation, we may use $M(-m^2_R)=M(-M^2_0)$. Therefore, the renormalized masses are 
$m_R=\pm M(-m_R)= E^{\pm}_R$, where $E^+_R=+|M(-m_R)|$ and $E^-_R=-|M(-m_R)|$ are positive and 
negative solutions, respectively. Using Eq.~(\ref{Mp}) with $m^2_R=M(-M^2_0)$ and $e^2=4\pi\alpha$, we find
\begin{equation}
\frac{|m_R|}{M_0}=1+\frac{\alpha}{2\pi} \int_0^1dx \frac{2+x}{\sqrt{1-x}}\ln\left[\sqrt{(1-x)\left(\frac{m^2}{M_0^2}-x\right)+x}\right].
\end{equation}
Considering $z_R\equiv m_R/M_0$ and $z\equiv m/M_0$, we have Eq.~(\ref{zR}) and Eq.~(\ref{fxz}). 
In the scalar-field case, given by Eq.~(\ref{scalar3D}), the critical point $z_c$ is the same. 
This is not surprising because of the similarities of the electron-self energy and the bosonic propagators.

\end{document}